# Edge-dislocation interactions with sub-3 nm vacancy and interstitial clusters in Fe–Cr alloys: Mechanisms, Obstacle strength and Fate

Layatmika Sahoo, Suchandrima Das*

*Department of Materials Engineering, Indian Institute of Science, Bengaluru 560012, India*

**suchandrima@iisc.ac.in*

## Highlights

- Sub-3 nm vacancy platelets are stronger than interstitial clusters and always survive
- Vacancy platelet strength follows strong-obstacle scaling, independent of Cr
- Interstitial clusters are mostly absorbed; strength varies fivefold with Cr layout
- Cr carried by interstitial loops collects at the loop–dislocation junction
- Computed α (0.02–0.25) indicates prior fitted α is inflated by TEM-invisible defects

## Abstract

Fe–Cr alloys harden at irradiation doses where transmission electron microscopy (TEM) shows no defects, implicating defects too small to be resolved. Their obstacle strength is unknown, so hardening models ignore them and fit measured hardening to visible defects alone. Here, molecular dynamics drives an edge dislocation through relaxed 1-3 nm sized vacancy and mixed Fe–Cr interstitial clusters, in Fe–3Cr, Fe–5Cr and Fe–10Cr at 300 K. Vacancy clusters relax into platelets whose strength rises with size to about 350 MPa at 2.9 nm, following strong-obstacle scaling independently of Cr content, and always survive dislocation passage. Interstitial clusters relax into ½⟨111⟩, mixed or ⟨100⟩ loops. They are mostly weaker than platelets of the same size, their strength varies up to fivefold with the local Cr arrangement, and most are absorbed. Only ⟨100⟩ loops in Fe–5Cr and Fe–10Cr often survive and during pinning, Cr collects at the junction between these loops and the dislocation; atom tracking shows that more than half comes from the loop itself. Surviving loops are not stronger than absorbed ones: strength and survival are thus distinct. Computed dispersed-barrier strengths ($\alpha \approx 0.02$–$0.25$) are upper bounds that lie below the α values obtained in earlier studies by fitting hardening to TEM-visible loops. These earlier values are therefore inflated by TEM-invisible defects. Wherever small vacancy clusters form, they are more effective obstacles than interstitial clusters of the same size. These defect-specific strengths and fates provide a physical basis for models of low-dose hardening.

**Keywords:** Fe–Cr alloys; irradiation hardening; molecular dynamics; dislocation loops; vacancy clusters; obstacle strength

## 1. Introduction

Ferritic and ferritic/martensitic (F/M) steels are leading candidate structural materials for fusion and advanced fission reactors [1–3]. Their body-centred cubic (bcc) structure gives them much better resistance to void swelling and irradiation creep than austenitic stainless steels [4], together with high thermal conductivity and low thermal expansion [3,5]. In reduced-activation (RAFM) grades such as F82H and EUROFER, elements that dominate the induced radioactivity, such as Nb and Mo, are replaced by Ta and W, while Cr and V are retained [3,6,7]. In all of these steels, the matrix is an Fe–Cr solid solution.

Neutron irradiation creates vacancies and self-interstitial atoms (SIAs) in collision cascades [8,9]. Most of them recombine. The survivors, and small clusters formed directly in the cascade, migrate and aggregate into dislocation loops and vacancy clusters. These defects obstruct dislocation glide and cause irradiation hardening. Cr changes this process: in neutron-irradiated Fe–Cr, the hardening varies non-monotonically with Cr content, with a minimum near 9% Cr [10]. Binary Fe–Cr alloys are therefore used as model materials, free of the carbides, martensite and minor additions of engineering steels [10,11].

Transmission electron microscopy (TEM) is the main tool used to characterise this damage. In Fe and Fe–Cr irradiated with heavy ions, the first visible defects are ½⟨111⟩ and ⟨100⟩ loops 2–5 nm in size [12]. ⟨100⟩ loops are more common, particularly in pure Fe, and Cr reduces loop mobility, so that more loops are retained [12]. These loops grow from much smaller clusters that TEM cannot resolve. In Fe–Cr, no defects are visible by TEM below about $10^{-3}$ dpa [13]. TEM therefore misses the smallest, earliest part of the damage.

Several observations show that these invisible defects contribute to hardening. The clearest come from very low doses. Song et al. irradiated Fe and Fe–Cr alloys containing 3–12 wt.% Cr with 20 MeV Fe ions at 313 K, to doses from $8\times10^{-5}$ to 6 dpa [14]. Hardening of 2–5% was measured at $8\times10^{-5}$ dpa, about ten times below the dose at which TEM first shows defects [13,14]. At $8\times10^{-4}$ dpa, the measured lattice strain corresponds to a defect density of about $10^{25}$ $m^{-3}$, roughly four orders of magnitude above the density estimated by TEM at a similar dose [14]. Most of the defects present are therefore invisible to TEM, and the hardening and lattice strain at these doses must come from defects too small to be seen. Positron annihilation spectroscopy detects some of these sub-resolution defects directly: in neutron-irradiated Fe, vacancies form clusters, whereas in Fe–Cr irradiated at 300 °C they remain mostly single vacancies, because Cr hinders vacancy clustering [11].

At higher doses, TEM resolves loops, but they do not account for all of the hardening. Matijasevic and Almazouzi irradiated Fe–Cr alloys containing 2.5–12% Cr with neutrons at 300 °C to 0.06, 0.6 and 1.5 dpa, and chose the loop strength α in the dispersed-barrier model so that the TEM-visible loops reproduced the hardening at 0.06 dpa [10]. With the same α, the visible loops fell increasingly short at higher doses: their density stayed nearly constant, whereas the hardening kept rising. The authors attributed the difference to invisible defects that accumulate with dose [10]. In a re-analysis of the same data, α had to rise from 0.6 at 0.06 dpa to 0.9 at 0.6 dpa and above 1 at 1.5 dpa [15]. A value above 1 exceeds the Orowan limit, which no single obstacle can exceed. Invisible defects, strengthening of the loops by Cr segregation, or both, were proposed to explain the difference [15]. In alloys with 9% Cr or more, Cr-rich α′ particles of about 2 nm, detected by small-angle neutron scattering, add further hardening [16,17]. The invisible population also changes with dose: in Fe–Cr, the lattice strain, a measure of the total defect content, rises up to about 0.8 dpa and then decreases [14]. The part of the hardening that TEM cannot explain therefore differs from one irradiation to another. Invisible defects thus control the response at low dose and contribute at higher doses, yet their structure, density and strength cannot be measured directly.

The inability to characterise these defects presents a fundamental difficulty for modelling irradiation hardening. Most continuum descriptions, such as crystal plasticity (CP) and dislocation dynamics models, require the density, size and obstacle strength of the individual defect populations. However, when only TEM-visible defects are included, the contribution of unresolved defects is not explicitly represented. In the widely used dispersed-barrier model, each defect type contributes to hardening in proportion to its obstacle strength, α, and to the square root of its density and size [18]. α is meant to be a property of the defect: a loop of a given size should have the same α at any dose. In practice, α is obtained by fitting the measured hardening to the TEM-visible defects alone. The CP model of Patra and McDowell for irradiated bcc Fe, for example, uses a single loop strength calibrated to experiments [19]. Any hardening from invisible defects is then attributed to the visible ones.

Because the invisible contribution differs from one irradiation to another, so does the fitted strength. Two neutron-irradiation studies of similar Fe–Cr alloys, at 0.6 and about 1.8 dpa, measured similar hardening but different densities of visible loops [20,21]. The fitted loop strengths therefore differed by a factor of two: $\alpha = 0.44$ at 0.6 dpa and 0.22 at 1.8 dpa. Neither the hardening measurements nor the TEM need be in error. Each fit absorbed a different, unmeasured contribution from defects that TEM did not resolve. In these analyses, loops below about 3 nm were not counted [20], or invisible clusters were neglected because no model for their strength was available [22]. A fitted α is therefore not the strength of the defect it is assigned to. It mixes the strength of the visible defects with the effect of an unknown population of invisible ones, and the resulting picture of which defects cause hardening is distorted.

Such fits reproduce the data used to obtain them. The limitation of fitted obstacle strengths, however, extends beyond uncertainty in the magnitude of the hardening contribution. Such parameters also provide no information on how individual defects evolve during dislocation motion. This matters under conditions other than those used for fitting, such as a different temperature or cyclic loading, where the defect population evolves. Some defects are absorbed by dislocations and cleared from the slip planes. Others survive and continue to pin dislocations. Clearing of defects is seen experimentally as localised deformation: around nanoindents in Fe and Fe–10Cr irradiated to about 1 dpa, slip steps appear, pile-up increases and the plastic zone is confined [23], as also reported for other irradiated Fe–Cr alloys [24,25]. Predictive models therefore need obstacle strengths, and defect fates, calculated for each defect type, including those below the TEM limit.

Atomistic simulations can provide this information. Molecular dynamics (MD) follows a dislocation as it meets a single defect, giving the critical stress needed to pass it and showing whether the defect is absorbed, cut or survives. In Fe–Cr, MD shows that Cr in solid solution raises the glide stress of an edge dislocation, and that Cr precipitates and voids are stronger obstacles than in pure Fe, with a strength that increases with size [26]. The strength of Cr-rich precipitates also depends on temperature and composition [5]. Interstitial loops have been studied the most. A loop is either absorbed by the dislocation or released through the formation and closure of a screw dipole, depending on loop size, Burgers vector, temperature and dislocation velocity [15,27]. For a 3.5 nm $\frac{1}{2}\langle 111 \rangle$ loop in pure Fe at 300 K, the critical stress is about 330 MPa [15].

Segregated Cr strengthens loops. Terentyev and Bakaev placed Cr around loops in Fe–9 at.% Cr by Monte Carlo simulation at 600 K, giving 25–50 at.% Cr at the edge of the loop [27]. The segregated Cr prevented complete absorption of the loops and made loops of 2 nm and below stronger obstacles. Without segregated Cr, loops of about 1 nm were absorbed with little resistance [27]. Cr and Al have similar effects in Fe–Cr–Al [28]. Vacancy-type defects have received less attention. In Fe–Cr, they have been represented only as spherical voids, which an edge dislocation shears at a stress that increases with void size [26]. Korchuganov et al. found that the glide stress in Fe–10Cr rises with the amount of radiation damage near a free surface, but did not isolate the contribution of individual vacancy clusters [29]. To our knowledge, vacancy clusters below the TEM limit have not been studied as obstacles in Fe–Cr.

These studies leave three gaps. First, the defects are idealised: loops are assumed to be circular, with a chosen Burgers vector and habit plane, and voids to be spherical. Clusters below the TEM limit contain only tens of point defects. After relaxation they need not take these shapes, and their strength depends on the shape they actually adopt. Second, Cr has been included only as precipitates [5,26] or as an equilibrium atmosphere around the defect [27,28]. Forming such an atmosphere requires Cr diffusion, which is slow near room temperature, so at low dose defects meet dislocations while Cr is still randomly distributed. This state has not been studied. Third, the only evidence that very small loops harden Fe–Cr comes from the segregated case [27], so the available MD cannot explain the hardening measured at $8\times10^{-5}$ dpa and 313 K [14]. There are therefore no atomistic data on the strength of relaxed vacancy and interstitial clusters smaller than 3 nm across a range of Cr contents. These are the data needed to give CP and dislocation dynamics models of low-dose Fe–Cr a physical basis.

This work addresses these gaps. We use MD to drive an edge dislocation through vacancy and interstitial clusters, 1–3 nm in size, in Fe–3Cr, Fe–5Cr and Fe–10Cr at 300 K, with Cr distributed randomly. Vacancy clusters are made by removing neighbouring atoms, and interstitial clusters from mixed Fe–Cr ⟨110⟩ dumbbells, the most stable interstitial configuration in Fe–Cr [30]. Each cluster is relaxed before the dislocation arrives, so its structure is a result of the simulation rather than an assumption. Both defect types are studied because it is not known which dominates the early damage [12,13]. The compositions match model alloys studied by nanoindentation and Laue microdiffraction after low-dose self-ion irradiation [14,23,31,32]. Obstacle strengths are reported as critical resolved shear stresses, which CP and dislocation dynamics models can use directly.

We use an edge dislocation for three reasons. First, it allows direct comparison with earlier MD studies of obstacle strength in Fe and Fe–Cr [5,15,26,27]. Second, bcc screw dislocations need thermal activation to move, and at MD strain rates their lattice resistance would mask the contribution of the obstacle [33]. Third, beneath an indenter in a bcc metal, dislocation loops expand mainly by the motion of their edge segments [34], which therefore sweep the irradiated layer in the nanoindentation experiments on these alloys [14,23,25,31,32]. TEM also shows small loops collecting along non-screw dislocations in irradiated Fe–3Cr, Fe–5Cr and Fe–8Cr at 250–350 °C [35]. Screw dislocations may interact differently with loops and are left for future work.

We address three questions. (i) How do cluster size and Cr content determine the relaxed structure of the clusters? (ii) Which interaction mechanisms do these structures produce with a moving edge dislocation? (iii) How do the obstacle strength, and the survival of the defect after the dislocation has passed, depend on defect type, size and Cr content? We first measure the glide resistance of the defect-free alloys, which provides the baseline for the obstacle strengths.

## 2. Methods

### 2.1 Simulation cell and alloys

Molecular dynamics simulations were performed using LAMMPS [36]. Atomic interactions were described by the two-band embedded-atom method potential of Bonny *et al.* [37]. Three alloys containing 3.21, 5.40 and 10.10 at.% Cr (hereafter Fe–3Cr, Fe–5Cr and Fe–10Cr) were constructed by randomly replacing Fe atoms in bcc cells with Cr. The lattice parameters, 2.8607, 2.8619 and 2.8651 Å for Fe–3Cr, Fe–5Cr and Fe–10Cr, respectively, were obtained by relaxing each alloy at 300 K with this potential. The same random seed was used for Cr substitution in all main simulations, so that cases of the same composition share the same Cr arrangement away from the defect. The local Cr environment of each defect nevertheless differs between cluster sizes, and its influence is assessed by repeating key cases with different seeds (Section 3.4 and Supplementary Material).

An edge dislocation (Burgers vector ½[111] along X, line direction [11$\bar{2}$] along Y and slip plane ($\bar{1}$10), normal to Z) was introduced 100 Å from one edge of the cell (Fig. 1a; construction in Appendix A). Periodic boundary conditions were applied along X and Y [38]. The cell measures 54.0 × 10.5 × 27.1 $nm^3$ (1,317,555 atoms), giving a dislocation density of ~$7\times10^{14}$ $m^{-2}$, typical of plastically deformed metals [33]. Because of the periodicity along Y, each defect forms part of a row of identical obstacles 10.5 nm apart along the dislocation line, and the obstacle strengths reported correspond to this spacing.

## 2.2 Defect construction

Irradiation damage was represented by vacancy clusters and mixed Fe–Cr SIA clusters, with effective diameters of 1–3 nm. This range was chosen to cover clusters below and at the TEM visibility limit: the smallest loops resolved by TEM in irradiated Fe and Fe–Cr are 2–5 nm in size [12]. The size distribution of the smaller, invisible clusters has not been measured for the present alloys, so the range studied is an assumption, not a measured distribution. Voids and loop coalescence, which occur at higher doses, were not considered.

Vacancy clusters were built by selecting a seed atom on the intended glide path and removing it together with its nearest neighbours in order of increasing distance, until the target number of vacancies was reached; Fe and Cr atoms were removed as encountered. Mixed ⟨110⟩ Fe–Cr dumbbells were used for the SIAs because they have lower formation energies than Fe–Fe or Cr–Cr dumbbells [30].

For SIA clusters, the same nearest-neighbour procedure was first applied in a pure Fe crystal, and the selected Fe atoms were replaced by Cr. An additional Fe atom was then inserted 1.43 Å from each of these Cr atoms along −Z, forming a compact platelet of parallel ⟨110⟩ Fe–Cr dumbbells. The remaining Cr needed to reach the target composition was then distributed randomly with the fixed seed. This starting configuration is equivalent to a small faulted ⟨110⟩ interstitial platelet, which can unfault towards either ⟨111⟩ or ⟨100⟩ character on relaxation.

Each defect was introduced on the glide plane, 250 Å ahead of the dislocation along X (Fig. 1a). The cell containing both the dislocation and the defect was then energy-minimised by conjugate gradient and equilibrated at 300 K for 2 ns in the NVT ensemble before shearing.

All cases were first simulated with one Cr distribution (seed 1); these results are shown in the main figures. To assess the effect of the local Cr arrangement, the shear simulations of 50-, 70- and 100-SIA clusters in Fe–5Cr and Fe–10Cr were repeated for two further Cr distributions, generated with the same procedure and different random seeds (Supplementary Material). For these repeats, the defect-free seed-1 cell of each alloy was used as the reference. For an independently generated Cr distribution, the glide resistance of the defect-free alloys differed from seed 1 by about 13 MPa (Appendix C), well within the baseline fluctuation band (Section 3.4). In addition, the relaxed structure of 50-SIA clusters was determined for 15 Cr distributions in each of Fe–3Cr and Fe–10Cr, and that of 25-SIA clusters in Fe–3Cr, in a smaller cell without a dislocation (Section 3.2).

The size of each relaxed defect is characterised by an effective diameter, $d = 2(A/\pi)^{1/2}$, where A is the area of the two-dimensional convex hull of the defect atoms after projection onto their best-fit plane. The defect atoms (those not identified as bcc by polyhedral template matching (PTM) [39] in OVITO [40]) were exported for each relaxed defect, the best-fit plane was obtained from a singular value decomposition of their positions, and the convex hull was computed in that plane. This equivalent-circle diameter gives a consistent measure across the different morphologies; it does not correspond to the physical diameter of a circular loop for irregular or tangled structures.

### 2.3 Shear simulations and obstacle strength

A few atomic layers at the bottom and top of the cell were kept fixed, and the top layers were displaced along X at a constant shear strain rate of $10^{-5}$ ps$^{-1}$ ($10^{7}$ s$^{-1}$) at 300 K, driving the edge dislocation towards the defect. This strain rate lies in the regime where the obstacle strength is nearly rate-independent (Appendix B).

The interaction force between the sheared top boundary and the mobile atoms was computed with the LAMMPS group/group command. Its X-component, $F_x$, was recorded at each timestep and converted to the resolved shear stress $\tau = F_x/(L_xL_y)$, where $L_xL_y$ is the area of the sheared boundary.

Fig. 1b shows representative curves for defect-free Fe–3Cr and for Fe–3Cr containing a 100-SIA cluster. In the defect-free alloy, the stress fluctuates about a constant value, reflecting the intrinsic glide resistance of the alloy. With a defect, a pronounced peak appears when the dislocation is pinned. The obstacle strength, $\Delta\tau_c$, is defined as the difference between the peak stress during the interaction and the average stress of the defect-free alloy over the same interval, taken as the period over which the stress rises from the baseline, peaks and returns (shaded in Fig. 1b). Curves were block-averaged over 50 ps intervals, and the baseline was taken from the defect-free seed-1 cell of the same alloy (Section 2.2). Because the cell is periodic along X, the dislocation re-encounters the defect about 2000 ps after the first interaction; only the first encounter (peak before 2300 ps) is analysed.

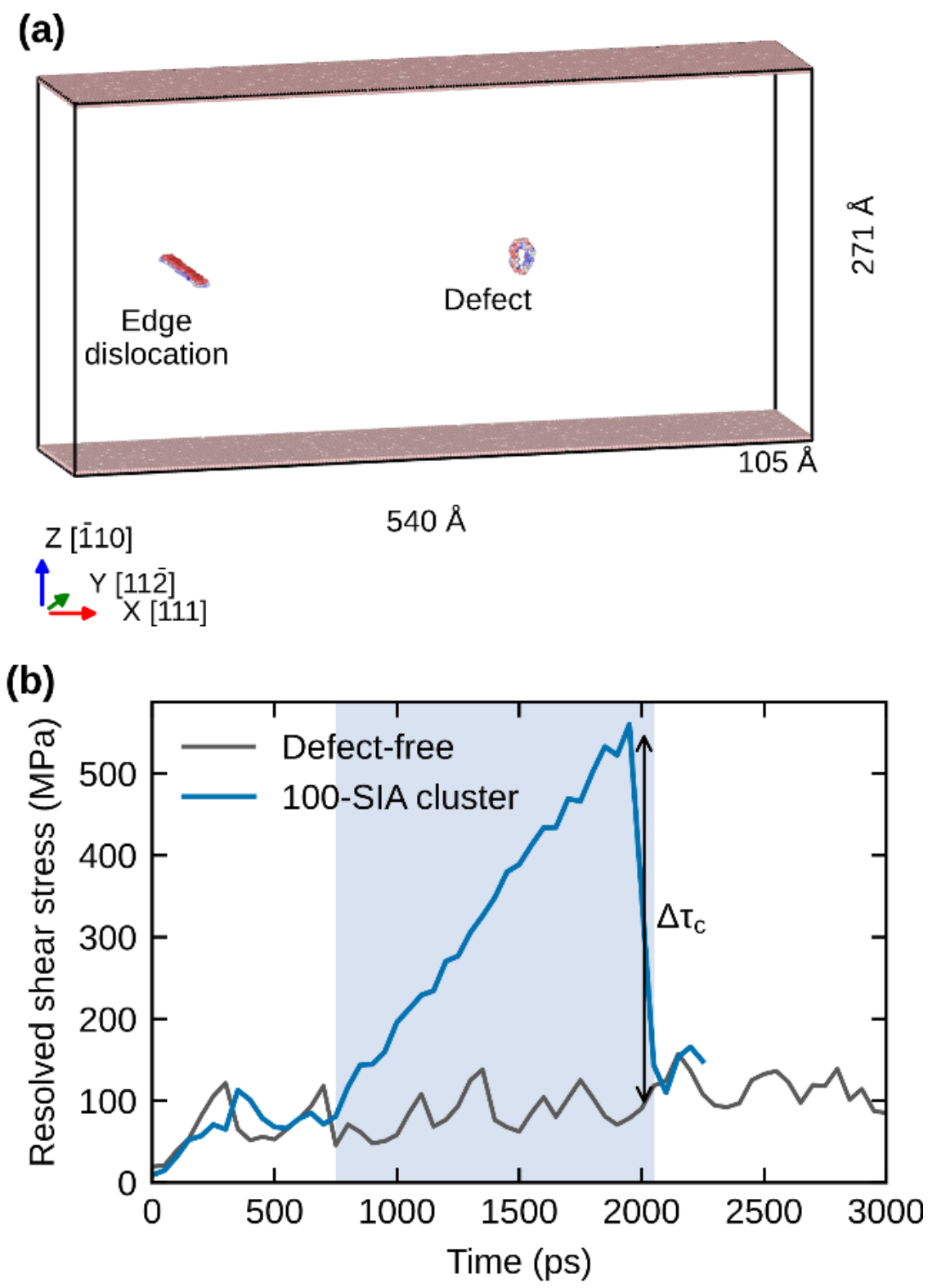

Figure 1 – (a) Simulation cell with an edge dislocation and a 100-SIA cluster in its path (Fe–3Cr). Atoms are coloured by the per-atom σxx·V (eV). (b) Resolved shear stress (50 ps block averages) as the dislocation moves through the cell at $10^{-5}$ $ps^{-1}$, for defect-free Fe–3Cr and Fe–3Cr containing the 100-SIA cluster; the shaded region marks the interaction interval.

# 3. Results

## 3.1 Glide resistance of the defect-free alloys

Fig. 2a shows the stress needed to sustain glide in the defect-free alloys, and Fig. 2b its average over the full simulation. The average glide resistance increases by ~22% from Fe–3Cr to Fe–5Cr and by a further 42% to Fe–10Cr, a total increase of ~73% (94, 114 and 163 MPa, respectively). Fe–10Cr also shows larger stress fluctuations, indicating a more heterogeneous local stress landscape. The increase is consistent with solid-solution strengthening by Cr, which modifies the local atomic environment of the core and raises the lattice friction [26]. Repeating these simulations with independently generated Cr distributions changes the absolute values only slightly and leaves the trend unchanged (Fig. 8, Appendix C), so the strengthening is set by the Cr concentration rather than by a particular configuration.

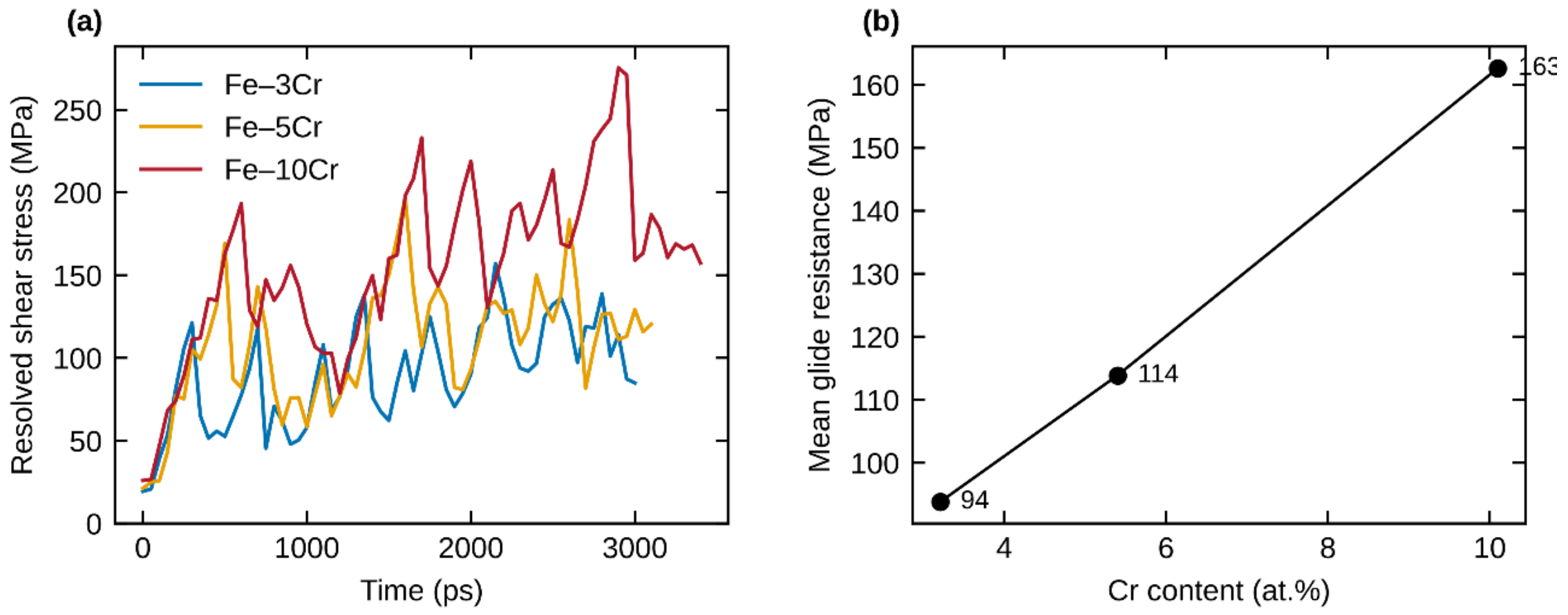

Figure 2 – (a) Resolved shear stress during glide in defect-free Fe–3Cr, Fe–5Cr and Fe–10Cr (50 ps block averages). (b) Average glide resistance versus Cr content over the complete simulation.

### 3.2 Relaxed structures of the defect clusters

***Interstitial clusters***

The relaxed loop character and effective diameter of each interstitial cluster are given in Table 1. In Fe–3Cr, the 10-SIA cluster relaxes into a ⟨111⟩ loop, the 20-SIA cluster into a ⟨100⟩ loop, and the 33-SIA cluster into a tangled configuration with ⟨111⟩ and ⟨100⟩ segments (referred to below as a mixed loop); from 50 SIAs upwards, ⟨100⟩ loops form. In Fe–5Cr, the 10- and 33-SIA clusters form ⟨111⟩ loops and the larger clusters ⟨100⟩ loops. In Fe–10Cr, ⟨111⟩ loops persist up to 70 SIAs and a ⟨100⟩ loop forms only at 100 SIAs. For this Cr distribution, ⟨100⟩ loops therefore appear at larger sizes in Fe–10Cr. With other Cr distributions, however, the 50- and 70-SIA clusters in Fe–10Cr also formed ⟨100⟩ or mixed loops (Supplementary Table S1), so the loop type at these sizes is not fixed by the Cr content alone.

To test how strongly the Cr arrangement controls the relaxed structure, clusters were relaxed for 15 independent Cr distributions in a smaller cell without a dislocation (205.6 × 203.2 × 202.2 Å$^3$, ~722,100 atoms). In Fe–3Cr, 25-SIA clusters gave 9 ½⟨111⟩ loops, 3 ⟨100⟩ loops and 3 unfaulted clusters, and 50-SIA clusters gave 11 ⟨100⟩ loops and 4 mixed structures. In Fe–10Cr, 50-SIA clusters gave 7 ⟨100⟩ loops, 4 ½⟨111⟩ loops and 4 mixed structures (Supplementary Table S2). Two conclusions follow. First, in both alloys small clusters favour ½⟨111⟩ loops and larger clusters ⟨100⟩ loops. Second, higher Cr content appears to make ½⟨111⟩ loops more likely at 50 SIAs (4 of 15 in Fe–10Cr, none in Fe–3Cr), but does not prevent ⟨100⟩ loops from forming. With 15 distributions per alloy, this difference is suggestive rather than conclusive (Fisher exact test, $p \approx 0.1$). The ½⟨111⟩ loop obtained for seed 1 in the Fe–10Cr shear cell is therefore one of the less frequent outcomes. For an individual cluster, the loop type is set by the local Cr arrangement.

Two findings from earlier work help to interpret these results. First, both ⟨111⟩ and ⟨100⟩ loops are observed in irradiated Fe and Fe–Cr, and the fraction of ⟨100⟩ loops depends on Cr content, dose and temperature [12,13]. Experiments disagree on the effect of Cr. In self-ion-irradiated Fe–Cr at 250 °C, the fraction of ⟨100⟩ loops fell from 54% in Fe–3Cr to 27% in Fe–8Cr [35], whereas proton irradiation of Fe–8Cr to Fe–18Cr at the same temperature gave a ratio independent of Cr content [41]. The present results, a modest shift towards ½⟨111⟩ loops with a strong dependence on the local Cr arrangement, are consistent with this divided picture. Second, in Fe, simple elastic arguments favour ½⟨111⟩ loops, and ⟨100⟩ loops become increasingly favourable only at temperatures above about 300 °C, owing to the softening of the elastic constants of Fe at high temperature [42]. In the present simulations, ⟨100⟩ loops appear at sizes as small as 50 SIAs (2.1 nm). This is smaller than these energetics would suggest. A likely reason is the way the clusters are built. Each cluster starts as a platelet of parallel ⟨110⟩ dumbbells, which is a faulted configuration. On relaxation it can unfault towards either ½⟨111⟩ or ⟨100⟩ character, and the starting geometry may favour ⟨100⟩. The size at which ⟨100⟩ loops first appear should therefore be regarded as specific to this construction. Two results do not depend on it: in all three alloys, small clusters relax mostly into ½⟨111⟩ loops, and at 50–70 SIAs the loop type depends on the local Cr arrangement.

| **Fe–3Cr: SIAs** | **Loop type** | **d (nm)** |
|---|---|---|
| 10 | ⟨111⟩ | 1.243 |
| 20 | ⟨100⟩ | 1.517 |
| 33 | Tangled ⟨111⟩/⟨100⟩ | 1.829 |
| 50 | ⟨100⟩ | 2.084 |
| 70 | ⟨100⟩ | 2.497 |
| 100 | ⟨100⟩ | 2.755 |
| **Fe–5Cr: SIAs** | **Loop type** | **d (nm)** |
| 10 | ⟨111⟩ | 1.104 |
| 33 | ⟨111⟩ | 1.795 |
| 50 | ⟨100⟩ | 2.115 |
| 70 | ⟨100⟩ | 2.551 |
| 100 | ⟨100⟩ | 2.842 |
| **Fe–10Cr: SIAs** | **Loop type** | **d (nm)** |
| 10 | ⟨111⟩ | 1.159 |
| 33 | ⟨111⟩ | 1.871 |
| 50 | ⟨111⟩ | 2.330 |
| 70 | ⟨111⟩ | 2.770 |
| 100 | ⟨100⟩ | 2.828 |

Table 1 – Relaxed interstitial loop character and effective diameters in Fe–3Cr, Fe–5Cr and Fe–10Cr, for the Cr distribution used in the main figures (seed 1). Results for other distributions are given in Supplementary Table S1.

***Vacancy clusters***

Vacancy clusters relax into open, platelet-like structures (Fig. 3a) at all sizes and compositions. Their effective diameters are almost independent of Cr content (Table 2). Earlier calculations for Fe provide the context [43,44]. For a vacancy cluster of any size, the lowest-energy form is a void. Planar vacancy clusters can therefore lower their energy by transforming into voids, but this transformation requires thermal activation, and planar clusters larger than about 2 nm are long-lived [43,44]. In the present simulations, no platelet transformed into a void or collapsed into a loop, either during the 2 ns relaxation or during the interaction. The platelets are therefore metastable structures. Whether they would transform during longer annealing cannot be determined on MD time scales.

| Vacancies | Fe–3Cr d (nm) | Fe–5Cr d (nm) | Fe–10Cr d (nm) |
|---|---|---|---|
| 10 | 0.941 | 0.939 | 0.950 |
| 33 | 1.416 | 1.417 | 1.417 |
| 40 | 1.625 | – | – |
| 50 | 1.713 | 1.711 | 1.726 |
| 70 | 1.980 | 1.982 | 1.977 |
| 100 | 2.263 | 2.264 | 2.267 |
| 120 | 2.468 | 2.462 | 2.459 |
| 170 | 2.857 | 2.852 | 2.858 |

Table 2 – Effective diameters of the relaxed vacancy clusters.

Four characteristic relaxed morphologies are therefore obtained: vacancy platelets, small ⟨111⟩ loops, mixed ⟨111⟩/⟨100⟩ loops and larger ⟨100⟩ loops. Their interactions with the dislocation are examined next.

### 3.3 Interaction mechanisms

Fig. 3 shows representative interaction sequences for the four morphologies. For the interstitial defects, the dislocation extraction algorithm (DXA) [45] identifies ⟨111⟩ (green) and ⟨100⟩ (pink) segments.

**Vacancy platelets.** The simplest interaction occurs with vacancy platelets and is the same in all three alloys (Fig. 3a–e). The dislocation bows around the platelet, cuts through it and glides on. The platelet survives but becomes smaller, because some of its vacancies are absorbed by the dislocation core as it cuts the platelet. DXA confirms this: the dislocation line is straight before it meets the platelet, but carries a series of small steps after passage. These steps are jogs formed by the absorbed vacancies. Absorption of vacancies by climb as the dislocation leaves the obstacle is a general feature of dislocation interactions with voids and vacancy clusters [46]. Apart from this loss of vacancies, the platelet keeps its structure.

**⟨111⟩ loops: direct absorption.** Small clusters that relax into ⟨111⟩ loops (Fig. 3f–i) keep their ⟨111⟩ character as the dislocation approaches and are then absorbed. After passage only a small residual interstitial cluster remains, with no loop detected by DXA. This mechanism is observed for 10 SIAs in Fe–3Cr, 10 and 33 in Fe–5Cr and 10–70 SIAs in Fe–10Cr (seed 1), i.e. all ⟨111⟩ loops.

**Mixed ⟨111⟩/⟨100⟩ loops: transformation before contact.** For the mixed loop (Fig. 3j–n), the stress field of the approaching dislocation transforms the loop into a predominantly ⟨111⟩ configuration before contact. A ⟨100⟩ segment reappears transiently during pinning (inset, Fig. 3l), after which the interaction proceeds as for the smaller ⟨111⟩ loops, leaving a small residual cluster. This pathway is observed only in Fe–3Cr, for the 33-SIA mixed loop and also for the 20-SIA ⟨100⟩ loop.

**⟨100⟩ loops in Fe–3Cr: transformation during interaction.** For ⟨100⟩ loops in Fe–3Cr (Fig. 3o–s), the transformation to ⟨111⟩ occurs progressively during the interaction itself rather than before contact. Isolated regions of the loop first acquire ⟨111⟩ character, and the transformation spreads as deformation continues. At the same time, as the dislocation attempts to unpin, Cr atoms become concentrated at the loop–dislocation interface (red atoms, insets in Fig. 3q,r). Once the ⟨100⟩→⟨111⟩ transformation is complete, the loop is absorbed, leaving a small residual cluster mostly of Cr (inset, Fig. 3s).

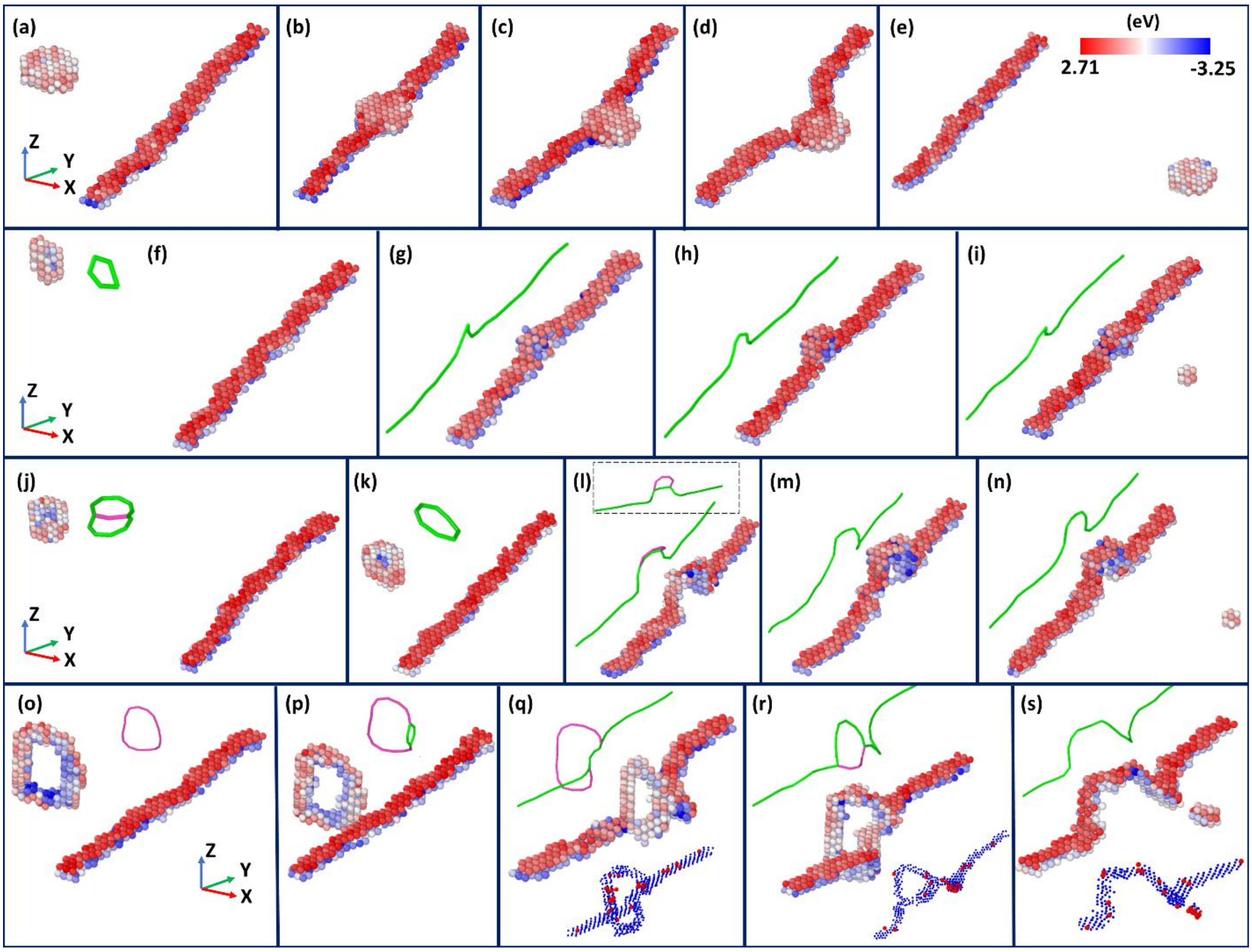

Figure 3 – Interaction of an edge dislocation with defect clusters in Fe–3Cr. (a)–(e) Vacancy platelet; (f)–(i) ⟨111⟩ loop (10 SIAs); (j)–(n) mixed ⟨111⟩/⟨100⟩ loop (33 SIAs); (o)–(s) ⟨100⟩ loop (100 SIAs). Atoms are coloured by per-atom $\sigma_{xx} \cdot V$ (eV). Insets: DXA ⟨111⟩ segments green and ⟨100⟩ segments pink; in (q)–(s) atoms are coloured by type, Cr (red) and Fe (blue). Axes as in (a).

**⟨100⟩ loops in Fe–5Cr and Fe–10Cr: pinning at a Cr-enriched junction.** A different pathway is found for ⟨100⟩ loops in Fe–5Cr and Fe–10Cr (Fig. 4). In the repeated simulations, all six 100-SIA ⟨100⟩ loops in these alloys left a residual defect, compared with three of six ⟨100⟩ loops of 50–70 SIAs; the others were fully absorbed. Initially, the Cr atoms of the mixed dumbbells are spread over the whole loop, and the matrix Cr is distributed randomly. When the dislocation is pinned, Cr accumulates at the loop–dislocation interface (Fig. 4h). Tracking the atom identities shows where this Cr comes from (Fig. 5). At peak pinning, 26 of the 50 Cr atoms of the original dumbbells are found at the junction, and they make up 26 of the 47 Cr atoms there. The other 21 are matrix Cr; 12 of these lie along the dislocation line away from the loop. The Cr carried by the loop is therefore drawn to the junction. This can happen on MD time scales because these Cr atoms are part of the interstitial loop, where atoms move far more easily than substitutional Cr in the matrix. Atom tracking was performed for this case only, but similar Cr enrichment at the junction is seen for the other ⟨100⟩ loops in Fe–5Cr and Fe–10Cr, and for the ⟨100⟩ loops in Fe–3Cr (Fig. 3q,r), which suggests the same origin. Experiments show a similar association between Cr and loops: Cr-enriched clusters form at dislocation loops in self-ion-irradiated Fe–5Cr and Fe–8Cr at 250 °C [35], and Cr segregates around loops in proton-irradiated Fe–Cr, which was attributed to Cr being dragged by the interstitial flux [41]. Structurally, the ⟨100⟩ loop becomes attached to a ⟨111⟩ dislocation segment, forming a twisted, connected junction that persists during pinning, and the Cr enrichment increases as it develops (rotated views, Fig. 4m,n; see also Fig. 9, Appendix D). The dislocation eventually unpins, leaving a smaller residual ⟨100⟩ loop with Cr concentrated at the region from which the dislocation disengaged (Fig. 4l).

This is consistent with earlier simulations showing that Cr enrichment can modify loop mobility and suppress complete loop absorption [15,27]. The contrast with the ⟨111⟩ loops in the same alloy (Fig. 3f–i), which are absorbed, shows that the role of Cr cannot be separated from the loop character: Cr enrichment at the junction occurs only for ⟨100⟩ loops, and loop survival only for ⟨100⟩ loops in Fe–5Cr and Fe–10Cr. TEM of irradiated Fe–Cr shows related behaviour: small loops collect along non-screw dislocations without merging with them, and often keep a Burgers vector different from that of the line [35].

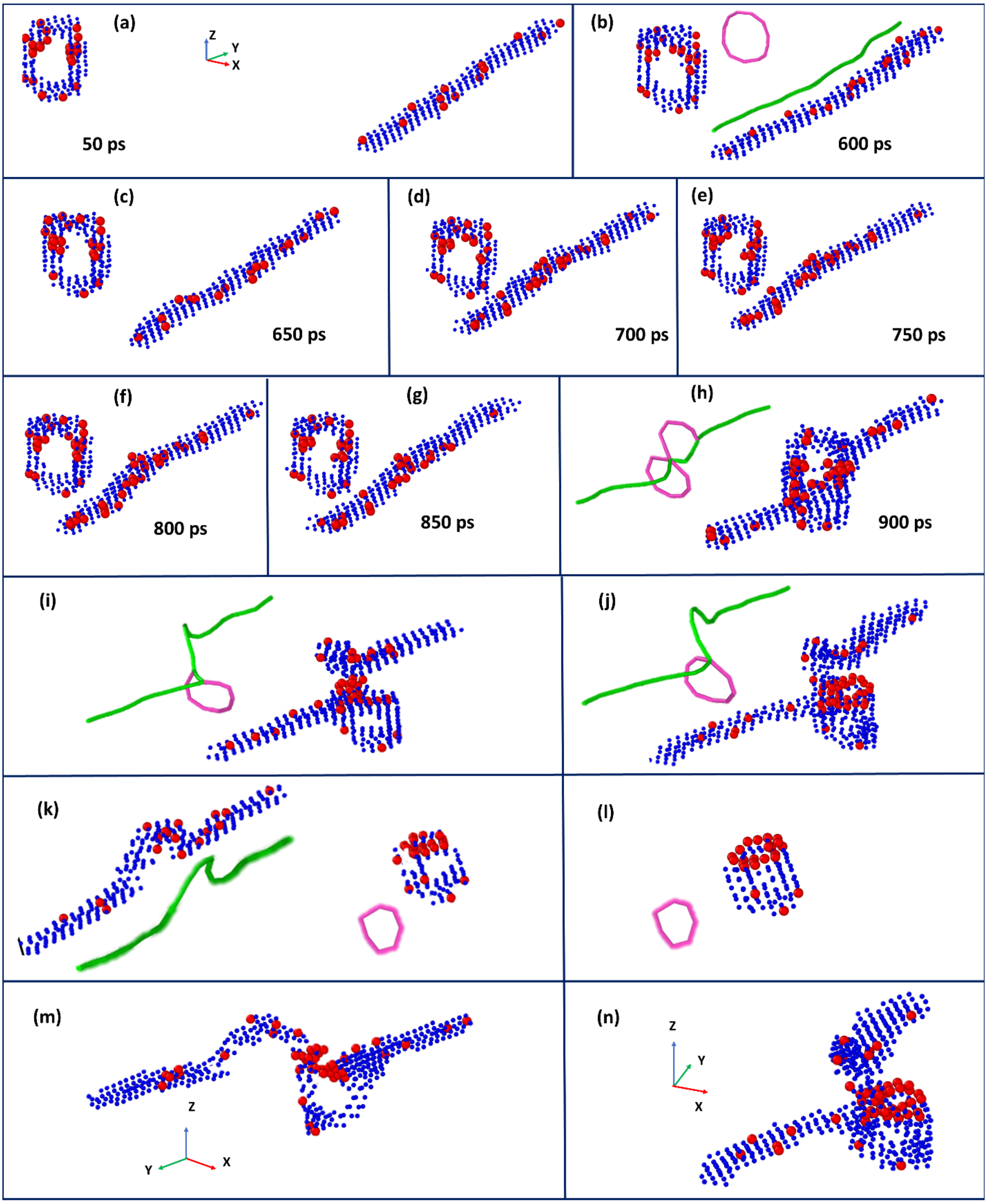

Figure 4 – Interaction of an edge dislocation with a ⟨100⟩ loop (50 SIAs) in Fe–5Cr. (a)–(k) Approach, formation of the connected junction during pinning, and separation. (l) Residual defect. (m),(n) Rotated views of (i) and (j) showing Cr enrichment at the junction. Atoms coloured by type: Cr red, Fe blue; DXA ⟨111⟩ (green) and ⟨100⟩ (pink) segments. Axes for (a)–(l) as in (a); (m),(n) have their own axes.

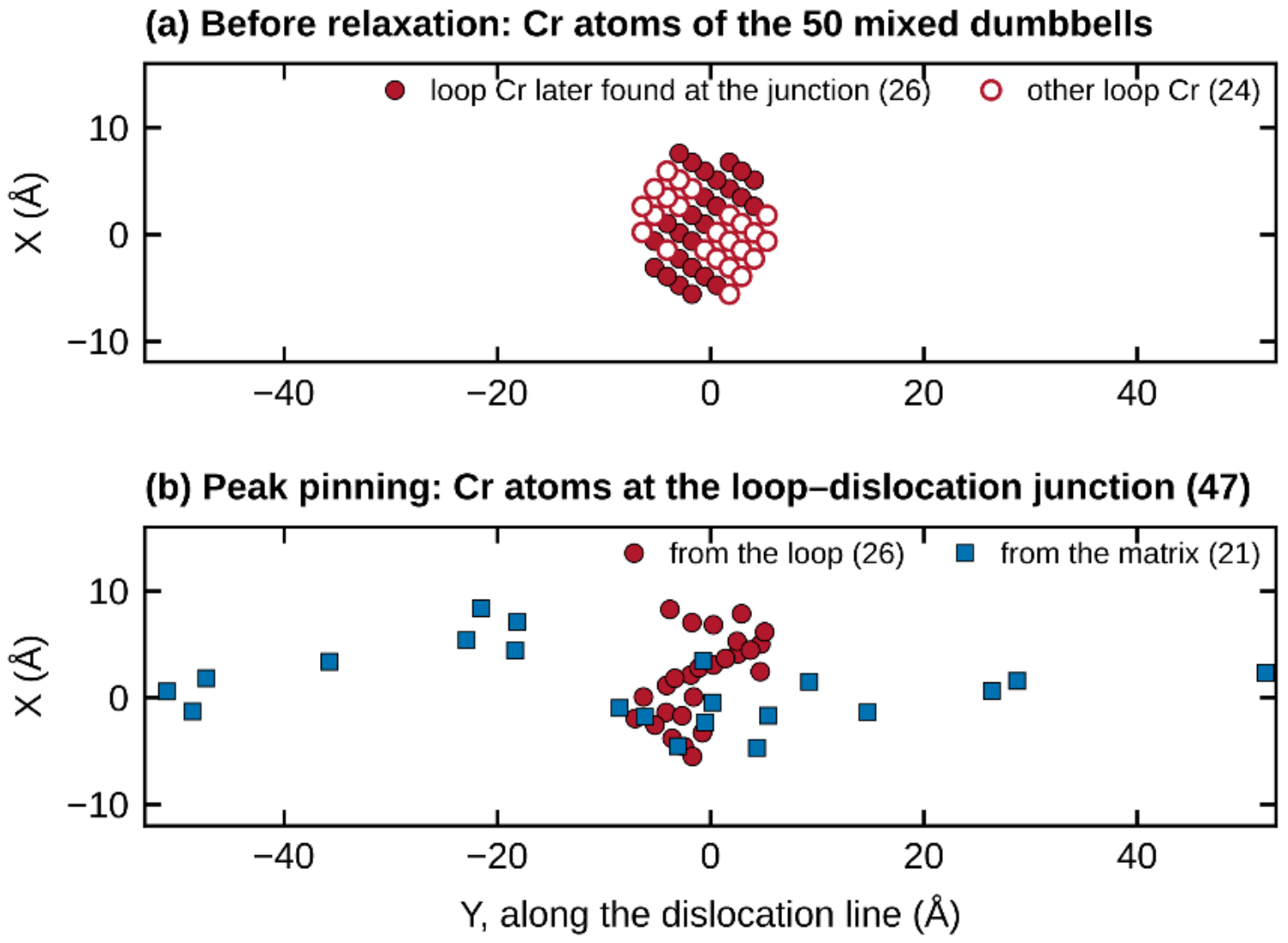

Figure 5 – Origin of the Cr atoms at the loop–dislocation junction for the 50-SIA ⟨100⟩ loop in Fe–5Cr of Fig. 4, projected onto the glide plane. (a) Cr atoms of the 50 mixed dumbbells before relaxation; filled symbols mark the 26 that are found at the junction at peak pinning. (b) The 47 Cr atoms at the junction at peak pinning, coloured by origin: from the loop (circles) or from the matrix (squares).

The interactions thus fall into four pathways (summarised in Table 3): (i) obstruction by vacancy platelets, which survive passage and remain obstacles to subsequent dislocations; (ii) direct absorption of ⟨111⟩ loops; (iii) transformation-mediated absorption of mixed and ⟨100⟩ loops in Fe–3Cr, with the transformation to ⟨111⟩ occurring before or during contact; and (iv) pinning of ⟨100⟩ loops at a Cr-enriched junction in Fe–5Cr and Fe–10Cr, with partial survival of a Cr-rich residual loop. Whether a defect obstructs, is absorbed or survives in a reconfigured state therefore depends on its relaxed character and, for ⟨100⟩ loops, on the alloy Cr content.

| **Pathway** | **Defect** | **Alloys / sizes** | **Defect after passage** |
|---|---|---|---|
| (i) Obstruction | Vacancy platelet | All | Smaller platelet |
| (ii) Direct absorption | ⟨111⟩ loop | 3Cr: 10; 5Cr: 10–33; 10Cr: 10–70 | Small SIA cluster |
| (iii) Transformation-mediated absorption | Mixed or ⟨100⟩ loop | 3Cr: 20–100 | Small (Cr-rich) cluster |
| (iv) Cr-enriched pinning | ⟨100⟩ loop | 5Cr, 10Cr: 50–100 (residual defect in 9 of 12 runs) | Smaller Cr-rich ⟨100⟩ loop |

Table 3 – Summary of interaction pathways.

### 3.4 Obstacle strength

Fig. 6 shows the obstacle strength, $\Delta\tau_c$, of each defect as a function of its size. The grey band marks the strength that the fluctuations of the defect-free alloys alone would give (65–85 MPa). A defect inside this band cannot be distinguished from the solid solution. Fig. 6 shows one Cr distribution (seed 1) for every case; the effect of the Cr distribution is described at the end of this section. Four results emerge.

**Vacancy platelets strengthen steadily with size, independently of Cr content** (Fig. 6a,c). The smallest platelets (about 1 nm) are weak obstacles, and the largest (2.5–2.9 nm) are strong ones ($\Delta\tau_c \approx$ 50–110 and 310–370 MPa, respectively). The trend is the same in all three alloys. A single curve, $\Delta\tau_c = A \ln(\bar{D}/b) + B$, describes the data of all three alloys ($R^2 = 0.83$, Fig. 6c). Here $\bar{D} = dL/(d + L)$ is the harmonic mean of the platelet diameter d and the obstacle spacing L. This is the size dependence derived by Bacon, Kocks and Scattergood for strong obstacles [47], and found in atomistic simulations of voids in Fe [46]. The fitted prefactor, A = 285 MPa, is close to the value $\mu b/(2\pi L)$ = 308 MPa expected for such obstacles. Vacancy platelets therefore behave as strong obstacles, and their strength is set by their size rather than by the Cr content.

**Interstitial clusters are mostly weaker than vacancy platelets** (Fig. 6b,d). Clusters of up to 33 SIAs (d $\lesssim$ 1.9 nm) are weak ($\Delta\tau_c \lesssim$ 150 MPa), and several lie within the noise band. Larger clusters (50–100 SIAs) give about 50–475 MPa, without a systematic dependence on size or Cr content. The two highest values (≈470 MPa) were obtained for the 70-SIA ½⟨111⟩ loop in Fe–10Cr and the 100-SIA ⟨100⟩ loop in Fe–3Cr.

**Loops that survive are not stronger than loops that are absorbed.** In Fe–5Cr and Fe–10Cr, ⟨100⟩ loops that left a residual defect gave 81–334 MPa, within the range of the clusters that were fully absorbed (93–468 MPa; all Cr distributions, Supplementary Table S1). Survival is therefore not a consequence of high strength. A likely reason lies in the interaction pathway: when the dislocation unpins by leaving part of the loop behind, the strength may be set by the local structure of the junction rather than by the whole loop.

**At the same size, vacancy platelets are stronger than interstitial clusters on average, and they survive.** For diameters of about 1.7–2.3 nm, vacancy platelets give 205–305 MPa in all three alloys, whereas interstitial clusters of similar size give mostly less than 150 MPa (Fig. 6c,d). Above about 2.5 nm, the platelets (310–370 MPa) remain stronger than most interstitial clusters, although individual interstitial loops can reach similar or higher values. Vacancy platelets always survived the passage of the dislocation, whereas interstitial clusters survived only as ⟨100⟩ loops, and not always. Obstacle strength and obstacle survival are therefore distinct properties.

**Effect of the Cr distribution.** For 50–100 SIAs in Fe–5Cr and Fe–10Cr, the shear simulations were repeated for two further Cr distributions (Supplementary Fig. S1 and Table S1). The relaxed clusters had similar sizes, within 0.3 nm of seed 1, but their strength varied by up to a factor of about five (97–468 MPa for 70 SIAs in Fe–10Cr). This variation is larger than the differences between sizes or alloys. The present data therefore show no systematic effect of Cr content on the strength of interstitial clusters. Part of the variation arises because the same cluster can relax into different loop types (Section 3.2). Large differences also occur between loops of the same type (109–277 MPa for 100-SIA ⟨100⟩ loops in Fe–5Cr). The repeats do not change the four results above. For every Cr distribution, interstitial clusters were on average weaker than vacancy platelets of the same size, ½⟨111⟩ and mixed loops were absorbed, only ⟨100⟩ loops survived, and surviving loops were not stronger than absorbed ones. The seed-1 values of 468 MPa (Fe–10Cr, 70 SIAs) and 476 MPa (Fe–3Cr, 100 SIAs) mark the upper end of the range; the repeats for Fe–10Cr at 70 SIAs gave 97 and 106 MPa.

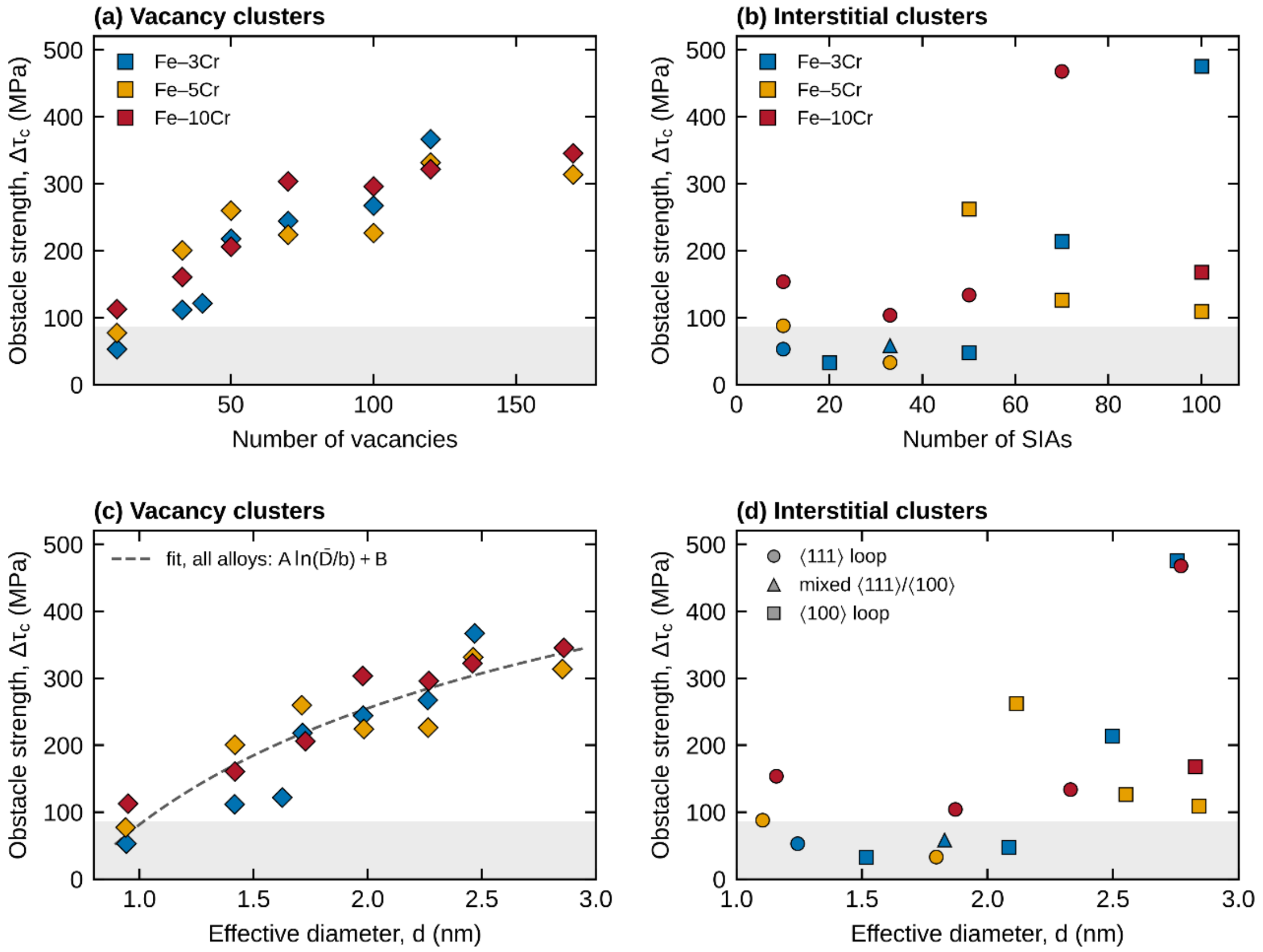

Figure 6 – Obstacle strength $\Delta\tau_c$ versus number of defects for (a) vacancy clusters and (b) interstitial clusters, and versus effective diameter for (c) vacancy clusters and (d) interstitial clusters in Fe–3Cr, Fe–5Cr and Fe–10Cr, for one Cr distribution (seed 1). Symbols indicate loop character (circle ½⟨111⟩, triangle mixed, square ⟨100⟩; diamonds vacancy platelets). The dashed line in (c) is a fit of $\Delta\tau_c = A\ \ln(\bar{D}/b) + B$ to the vacancy data of all three alloys, with $\bar{D} = dL/(d + L)$. The grey band marks the baseline fluctuation of the defect-free alloys. Results for other Cr distributions are given in Supplementary Fig. S1.

# 4. Discussion

## 4.1 Which defects can cause hardening at low dose?

The experiments reviewed in Section 1 show hardening of Fe–Cr at doses where no defects are visible by TEM [14]. The present results help to identify which of the invisible defects could cause it. Interstitial clusters with randomly distributed Cr are weak obstacles below about 2 nm and, on average, moderate obstacles up to 2.8 nm, and most of them are absorbed by the first dislocation. In this state, small interstitial clusters are therefore unlikely to cause significant hardening. They can become strong obstacles if Cr segregates to them [27], but segregation requires diffusion and is unlikely at low dose near room temperature. Vacancy platelets of the same size, in contrast, are on average about twice as strong, and they survive the passage of a dislocation. The same ranking was found in pure Fe, where a 37-SIA loop was a weaker obstacle than voids containing a similar number of vacancies [46]. Vacancies need not form compact voids to be effective: positron annihilation and MD show that cascades leave loose vacancy clusters, and such clusters are nearly as strong as voids with the same number of vacancies [46]. Per cluster, small vacancy clusters are therefore more effective obstacles than interstitial clusters. Whether they dominate the hardening measured below the TEM limit depends on how many of them form, which is less certain. Voids are not observed in irradiated Fe–Cr, in contrast to pure Fe [22,35], and positron annihilation of Fe–Cr neutron-irradiated at 300 °C to 0.06–0.6 dpa detects mainly single vacancies, whereas pure Fe irradiated to similar doses contains vacancy clusters [11]. Cr therefore appears to hinder vacancy clustering, at least at 300 °C. The vacancy population at the lower temperatures of the low-dose experiments, such as 313 K [14], has not been characterised. Wherever clusters of 10–170 vacancies form, the present results show that they are more effective obstacles than interstitial clusters of the same size.

This conclusion has a consequence for the effect of Cr. At low dose, the measured hardening increases with Cr content [14], although at higher doses the effect of Cr is not monotonic [32,35]. The present strengths of vacancy platelets, however, depend little on Cr. The effect of Cr would then have to act mainly through the defect population, rather than through the strength of individual defects. One possible route is the reduced mobility of small clusters in Fe–Cr [12,22], which would retain more clusters in the matrix. Testing this requires simulations of defect accumulation, which are beyond the present work.

The weakness of small interstitial clusters is itself an important result. The present clusters contain Cr, one atom in each mixed dumbbell, but no additional Cr segregated around them. Earlier MD showed that sub-2 nm loops become strong obstacles when Cr is first segregated to their perimeter, by Monte Carlo simulation at 600 K [27]. The present results show that, without such segregation, loops of this size contribute little. During pinning, Cr carried by a larger ⟨100⟩ loop does collect at the junction (Section 3.3), but this does not make the loop stronger (Section 3.4). The contribution of small interstitial clusters to hardening therefore depends on the irradiation temperature and dose, through the extent of Cr segregation before the interaction.

### 4.2 Comparison with earlier simulations

The interstitial loops of 2.1–2.8 nm studied here gave mean strengths of about 150–220 MPa at a spacing of 10.5 nm, with individual values up to about 470 MPa. Terentyev *et al.* obtained about 330 MPa for a 3.5 nm ½⟨111⟩ loop in pure Fe at 300 K and a spacing of about 21 nm [15]. Scaled to the present spacing by the approximately inverse dependence of strength on spacing, their value corresponds to roughly 600 MPa. This is above most of the present values, as expected for a larger loop, and the strongest present loops (≈470 MPa) approach it.

The ⟨100⟩ loops in Fe–5Cr and Fe–10Cr were only partly absorbed. Terentyev and co-workers found partial absorption for Cr-decorated loops as well [15,27]. In their simulations, however, Cr had been segregated to the loop by Monte Carlo simulation before the interaction, and the decorated loops were stronger than undecorated ones. Here, Cr gathers at the junction only during pinning, and loops that partly survive are not stronger than loops that are absorbed. Cr segregated before the interaction and Cr gathered during it therefore appear to have different effects on strength, although both lead to partial survival of the loop.

For vacancy defects, earlier MD in Fe–Cr considered spherical voids, whose strength increases with size [26]. The present vacancy clusters are planar, not spherical, but their strength shows the same increase with size. In atomistic simulations of voids and precipitates, the solid solution of Fe–10Cr was found to add an approximately constant stress, independent of obstacle size [46]. Subtracting the glide resistance of the defect-free alloy, as done here, therefore isolates the contribution of the obstacle.

### 4.3 Implications for crystal plasticity and dislocation dynamics models

The present strengths can be expressed in the form used by dispersed-barrier models. For an obstacle spacing L, the obstacle-strength parameter is $\alpha = \Delta\tau_c L/(\mu b)$. With L = 10.5 nm, b = 0.248 nm and $\mu \approx 82$ GPa for Fe, interstitial clusters give $\alpha \approx 0.02$–0.25 and vacancy platelets $\alpha \approx 0.03$–0.19. These values lie at or below the lower end of the α values obtained in earlier studies by fitting measured hardening to TEM-visible loops. For TEM-visible loops in neutron-irradiated Fe–Cr, fits gave α = 0.22 (range 0.18–0.26) [20] and 0.44 [21], and α = 0.4 has been assumed in hardening analyses [22] and in a CP model of irradiated Fe [19]. Values of 0.5 to above 1 were needed to explain the hardening of neutron-irradiated Fe–Cr by the visible loops alone [10,15]. The earlier fitted values are therefore inflated. Because the fits included only the TEM-visible loops, the hardening caused by invisible defects was attributed to those loops, which made them appear stronger than they are. The inflation is in fact larger than these numbers suggest. The present values are strengths of single obstacles in a regular row on the glide plane. A random array of obstacles of the same strength is easier to pass: for weak obstacles, the effective strength is approximately $(0.8 + 0.2\varphi_{c}/\pi)\alpha^{3/2}$, where $\varphi_{c}$ is the critical angle at which the dislocation breaks free [20]. For α = 0.25 ($\varphi_{c} \approx 150°$), this gives about 0.12. Obstacles away from the glide plane and the high strain rate act in the same direction, so the present α values are upper bounds for real microstructures. Because the vacancy platelets follow the $\ln(\bar{D}/b)$ dependence of strong obstacles (Section 3.4), their strength at other spacings can be estimated from the fit; for widely spaced obstacles, the α values would be about 10% larger, which does not change these comparisons.

Two consequences follow. First, a single fitted α cannot represent a defect population whose individual strengths differ by a factor of about ten. The dependence of the fitted values on the irradiation conditions (Section 1) points the same way. Second, the size of the correction can be estimated. The invisible clusters that the earlier analyses did not count [20,22] must supply the difference. Because hardening scales as $\alpha(Nd)^{1/2}$, a true strength of about 0.1 instead of a fitted 0.4 requires about 16 times more obstacles, Nd, to give the same hardening. Models should therefore describe each defect type separately, with its own strength and its own fate. Vacancy platelets should be treated as obstacles that persist. Most interstitial clusters should be treated as obstacles that are removed by the first dislocation. ⟨100⟩ loops in Fe–5Cr and Fe–10Cr should be treated as obstacles that often survive. This distinction controls how the obstacle population evolves under repeated or cyclic loading, and hence when defect-free channels form.

### 4.4 Limitations

The present results rest on several assumptions, which also point to directions for further work. Only edge dislocations were considered; screw dislocations, which can form helical turns with loops, may interact differently. Each defect forms part of a regular row of obstacles 10.5 nm apart on the glide plane; the strengths reported correspond to this configuration, and Section 4.3 discusses how they translate to random arrays. The shear rate of $10^7$ $s^{-1}$, typical of MD, lies in the range where the obstacle strength changes little with rate (Appendix B). Cr is distributed randomly, which represents the state at low doses near room temperature; segregation during longer irradiations at higher temperatures would add to the effects described here. The interstitial clusters were built from platelets of mixed ⟨110⟩ dumbbells, which sets their initial Cr content, and the size at which ⟨100⟩ loops first appear may depend on this construction. The repeated Cr distributions for Fe–5Cr and Fe–10Cr, and the relaxations for 15 distributions, quantify the role of the local Cr arrangement; the vacancy trends, smooth and consistent across three alloys, indicate that vacancy platelets are less sensitive to it. Finally, as in all MD studies, absolute values depend on the interatomic potential [37], whereas the trends with defect type, size and Cr content are expected to be robust.

## 5. Conclusions

MD simulations of edge dislocations interacting with relaxed vacancy and mixed Fe–Cr interstitial clusters in Fe–3Cr, Fe–5Cr and Fe–10Cr at 300 K lead to the following conclusions.

1. The glide resistance of the defect-free alloys increases from 94 MPa in Fe–3Cr to 114 MPa in Fe–5Cr and 163 MPa in Fe–10Cr (~73%), independently of the particular Cr distribution.
2. Vacancy clusters relax into open, metastable platelets. Interstitial clusters built from mixed dumbbells relax mostly into ½⟨111⟩ loops at small sizes, and into ½⟨111⟩, mixed or ⟨100⟩ loops at 50–100 SIAs, depending on size and on the local Cr arrangement. At 50 SIAs, ½⟨111⟩ loops formed in 4 of 15 Cr distributions in Fe–10Cr and in none in Fe–3Cr, but the type of an individual loop is set by the local Cr arrangement.
3. Four interaction pathways are identified: obstruction by vacancy platelets, which survive; direct absorption of ½⟨111⟩ loops; transformation-mediated absorption of mixed and ⟨100⟩ loops in Fe–3Cr; and pinning of ⟨100⟩ loops at a Cr-enriched junction in Fe–5Cr and Fe–10Cr, after which the loop often survives. During pinning, Cr collects at the junction of ⟨100⟩ loops; atom tracking for one loop shows that more than half of it comes from the loop itself.
4. Vacancy platelets strengthen steadily with size (50–370 MPa), following the $\ln(\bar{D}/b)$ scaling of strong obstacles independently of Cr content, and always survive. Interstitial clusters are mostly weaker (below 150 MPa up to 33 SIAs), their strength varies by up to a factor of five between Cr arrangements, and most are absorbed.
5. Strength and survival are distinct: loops that survive are not stronger than loops that are absorbed. The computed strengths, $\alpha \approx 0.02$–$0.25$, are upper bounds for random arrays of obstacles and lie below the α values obtained in earlier studies by fitting measured hardening to TEM-visible loops. Those earlier values are therefore inflated: they attribute to the visible loops the hardening caused by defects that TEM cannot resolve, and they are not transferable between irradiations. Wherever small vacancy clusters form, they are likely to contribute strongly to this hardening. CP and dislocation dynamics models of low-dose Fe–Cr should describe each defect type with its own calculated strength and fate, rather than with a single fitted α.

## CRediT authorship contribution statement

Layatmika Sahoo: Investigation, Formal analysis, Visualisation, Writing – original draft. Suchandrima Das: Conceptualisation, Supervision, Funding acquisition, Writing – review & editing.

## Declaration of competing interest

The authors declare that they have no known competing financial interests or personal relationships that could have appeared to influence the work reported in this paper.

## Data availability

Data will be made available on request.

## Declaration of generative AI and AI-assisted technologies in the writing process

During the preparation of this work the authors used Claude (Anthropic) to improve the language and organisation of the manuscript and to assist with figure preparation. After using this tool, the authors reviewed and edited the content as needed and take full responsibility for the content of the publication.

## Acknowledgements

L. Sahoo acknowledges the Ministry of Education (MoE), Government of India, for stipend support. S. Das acknowledges the Indian Institute of Science for a start-up research grant, which funded the high-performance computing facility used for these simulations.

## Appendix A. Construction of the simulation cell

The cell with an edge dislocation and periodic boundaries along X and Y was constructed as follows. (1) A cell of double the intended height (540.1 × 105.1 × 542.2 $Å^3$) was created. (2) Two edge dislocations along Y, at the mid-point in X, were introduced at Z = −135 Å and Z = 135 Å. (3) The cell was energy-minimised with periodic boundaries in all directions. (4) Only the atoms with Z > 0 were retained. (5) The resulting 540.1 × 105.1 × 271.1 $Å^3$ cell was minimised again with periodic boundaries along X and Y (Fig. 1a).

## Appendix B. Strain-rate sensitivity

Shear simulations of defect-free Fe–3Cr, and of Fe–3Cr containing a 70-SIA cluster, were performed at shear rates of $10^{-6}$, $10^{-5}$, $10^{-4}$ and $10^{-2}$ $ps^{-1}$ (Fig. 7). The mean glide resistance after 1000 ps changes strongly between $10^{-4}$ and $10^{-5}$ $ps^{-1}$ (from about 370 to 85 MPa) but little between $10^{-5}$ and $10^{-6}$ $ps^{-1}$ (85 and 77 MPa, a change of about 10%; Fig. 7a). The obstacle strength of the 70-SIA cluster, obtained with the defect-free baseline at the same rate, behaves in the same way (about 660, 214 and 206 MPa at $10^{-4}$, $10^{-5}$ and $10^{-6}$ $ps^{-1}$; a change of about 4% between the two lowest rates; Fig. 7b). At $10^{-2}$ $ps^{-1}$ both values are lower than at $10^{-4}$ $ps^{-1}$; at such rates the response is not quasi-static. A shear rate of $10^{-5}$ $ps^{-1}$ was therefore used as a compromise between rate independence and computational cost.

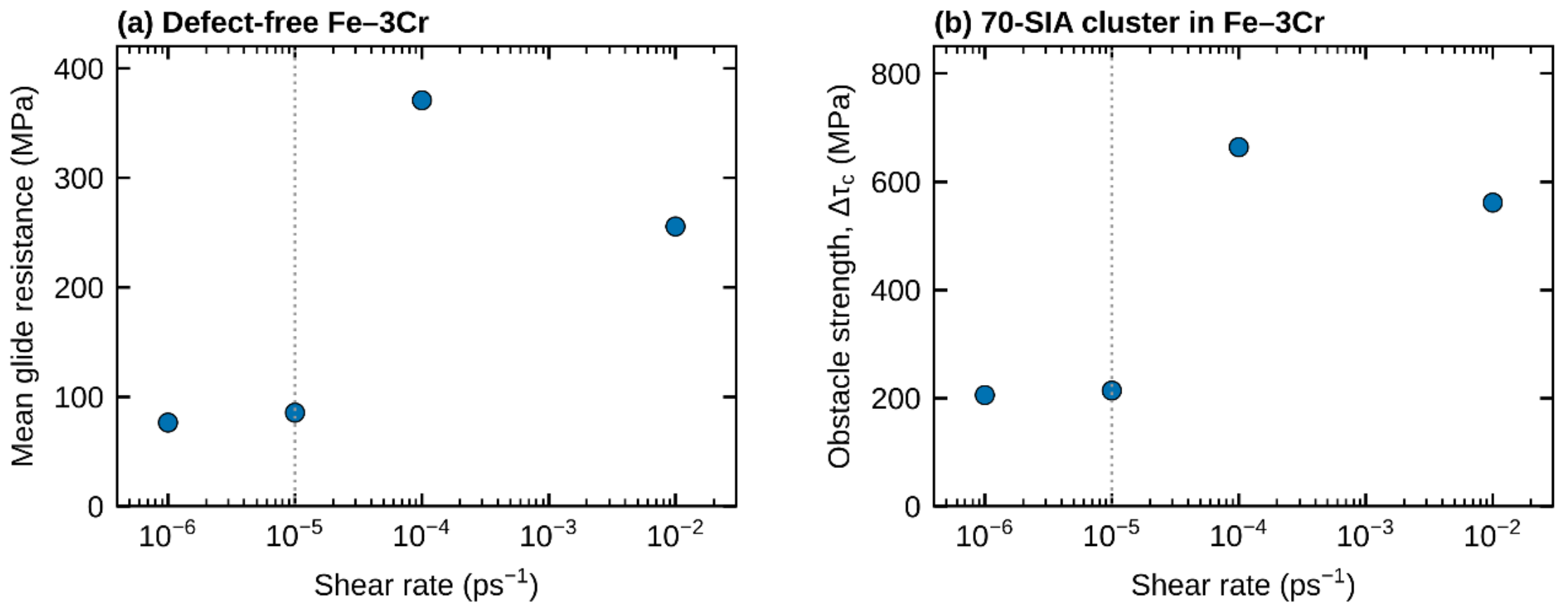

Figure 7 – Effect of shear rate in Fe–3Cr: (a) mean glide resistance of the defect-free alloy after 1000 ps and (b) obstacle strength of a 70-SIA cluster. The dotted line marks the rate used in this work ($10^{-5}$ $ps^{-1}$).

## Appendix C. Sensitivity to the Cr distribution

Fig. 8 compares the mean glide resistance of the defect-free alloys for the Cr distribution used in the main text (seed 1) and for an independently generated distribution. Both increase with Cr content, by 73% and 84% from Fe–3Cr to Fe–10Cr. The independent distribution gives values about 13 MPa lower at all three compositions. This difference is well within the baseline fluctuation band used for the obstacle strengths (65–85 MPa).

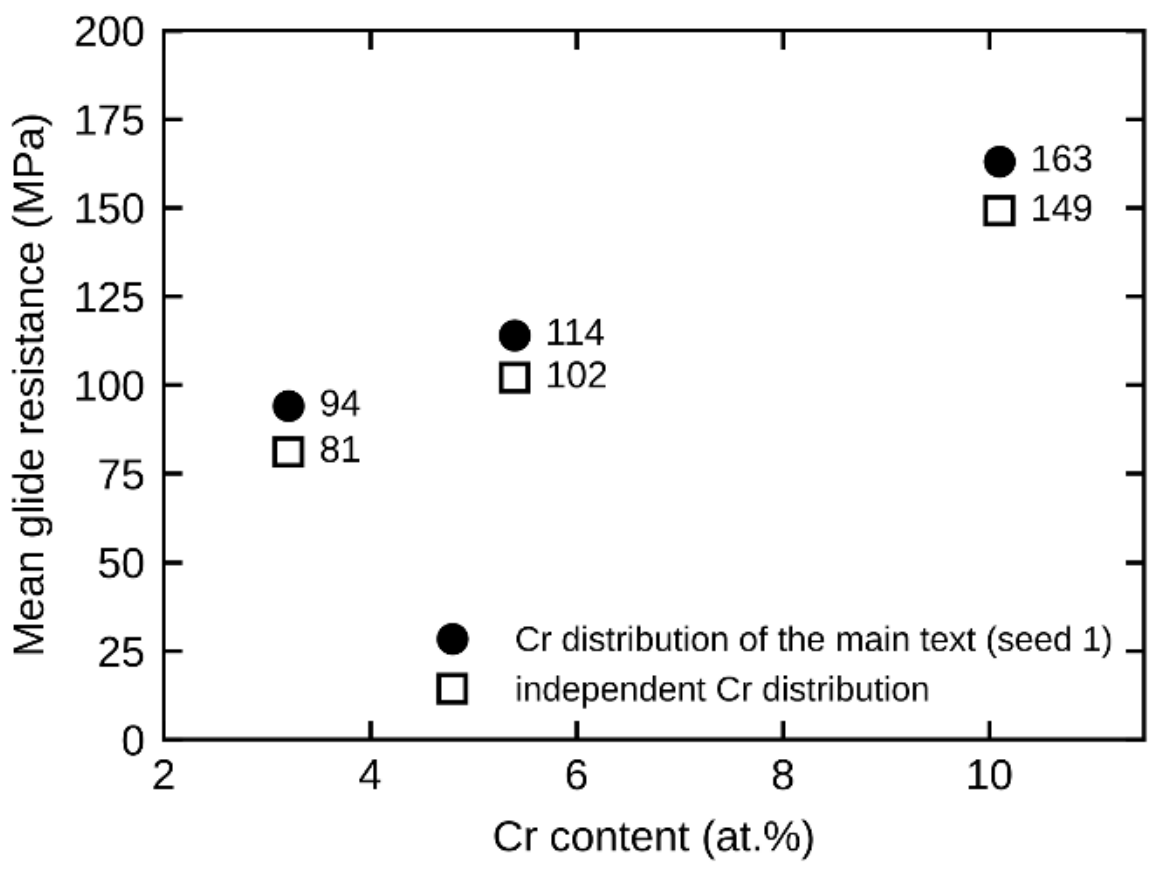

Figure 8 – Mean glide resistance of defect-free Fe–3Cr, Fe–5Cr and Fe–10Cr for the Cr distribution used in the main text (seed 1) and for an independently generated distribution (50 ps averages after 200 ps).

## Appendix D. Interaction with a 100-SIA ⟨100⟩ loop in Fe–5Cr

Fig. 9 shows the interaction of an edge dislocation with the 100-SIA ⟨100⟩ loop in Fe–5Cr. The loop follows the same pathway as the 50-SIA loop of Fig. 4: a connected junction forms during pinning, and Cr becomes concentrated at it.

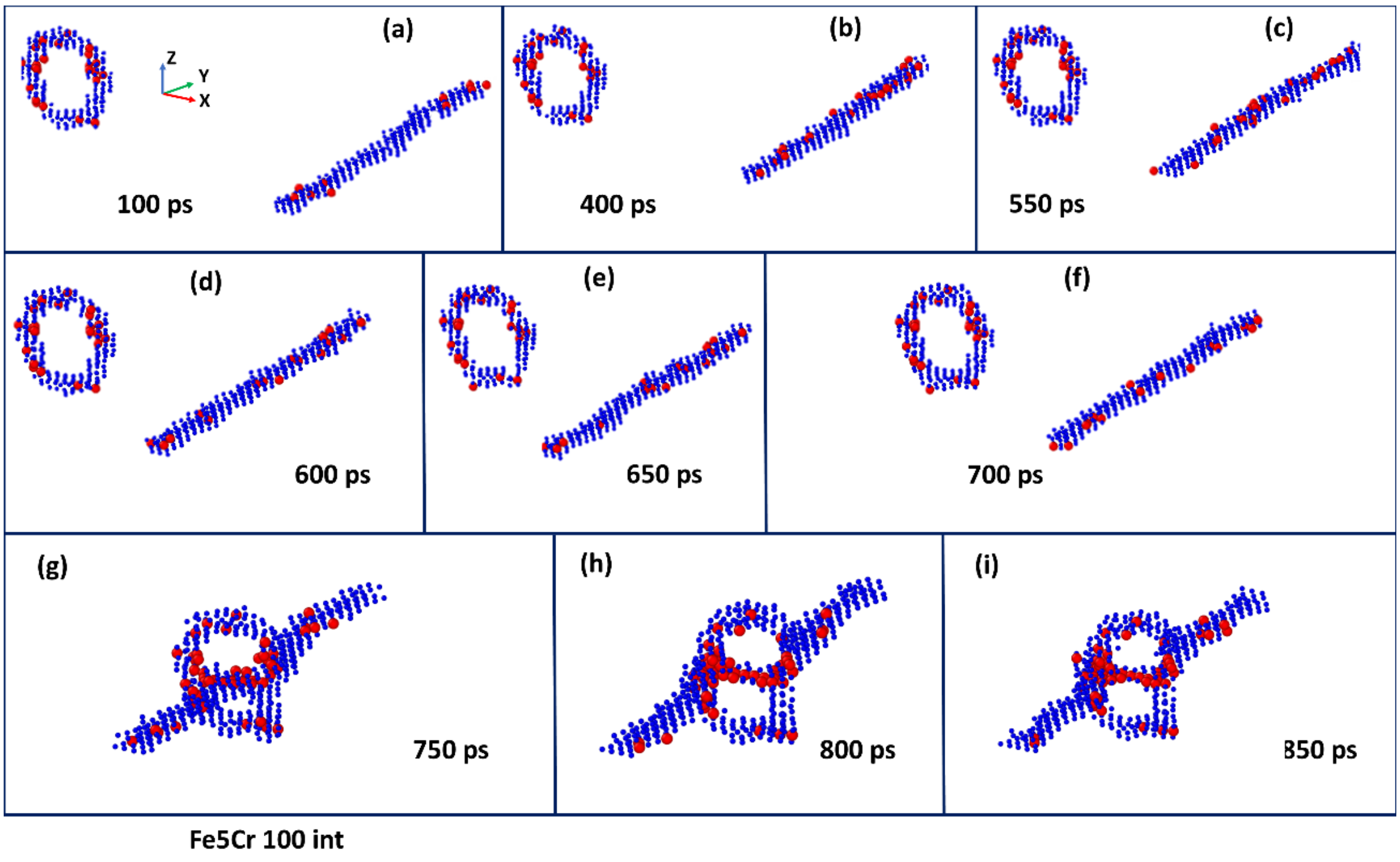

Figure 9 – Interaction of an edge dislocation with a ⟨100⟩ loop (100 SIAs) in Fe–5Cr. (a)–(i) Approach and formation of the connected junction during pinning. Atoms coloured by type: Cr red, Fe blue. Axes as in (a).

# Supplementary Material

***Edge-dislocation interactions with sub-3 nm vacancy and interstitial clusters in Fe–Cr alloys: Mechanisms, Obstacle strength and Fate***

Layatmika Sahoo, Suchandrima Das

Department of Materials Engineering, Indian Institute of Science, Bengaluru 560012, India

## S1. Repeated Cr distributions in the shear simulations

The shear simulations of 50-, 70- and 100-SIA clusters in Fe–5Cr and Fe–10Cr were repeated for two further Cr distributions (seeds 2 and 3), generated with the same procedure as seed 1 and different random seeds. The defect-free seed-1 cell of each alloy was used as the reference for the obstacle strength; for an independently generated Cr distribution, the glide resistance of the defect-free alloys differs from seed 1 by about 13 MPa (Appendix C of the main text), well within the baseline fluctuation band. Loop character was determined by DXA and the effective diameter as described in Section 2.2 of the main text. Table S1 lists the results; Fig. S1 compares them with seed 1, and Fig. S2 shows the stress–time curves. A residual defect counted as surviving in the main text is a ⟨100⟩ loop or tangled structure; a small cluster left after absorption is not.

Table S1 – Relaxed loop character, effective diameter d, obstacle strength Δτc and defect after passage for each Cr distribution.

| **Alloy** | **SIAs** | **Seed** | **Loop type** | **d (nm)** | **Δτc (MPa)** | **Defect after passage** |
|---|---|---|---|---|---|---|
| **Fe–5Cr** | 50 | 1 | ⟨100⟩ | 2.12 | 262 | residual ⟨100⟩ loop |
| **Fe–5Cr** | 50 | 2 | mixed ½⟨111⟩/⟨100⟩ | 2.15 | 110 | absorbed, small cluster left |
| **Fe–5Cr** | 50 | 3 | ⟨100⟩ | 2.07 | 135 | fully absorbed |
| **Fe–5Cr** | 70 | 1 | ⟨100⟩ | 2.55 | 126 | residual ⟨100⟩ loop |
| **Fe–5Cr** | 70 | 2 | ½⟨111⟩ | 2.55 | 93 | fully absorbed |
| **Fe–5Cr** | 70 | 3 | ½⟨111⟩ | 2.58 | 227 | absorbed, small cluster left |
| **Fe–5Cr** | 100 | 1 | ⟨100⟩ | 2.84 | 109 | residual ⟨100⟩ loop |
| **Fe–5Cr** | 100 | 2 | ⟨100⟩ | 2.91 | 169 | residual ⟨100⟩ loop |
| **Fe–5Cr** | 100 | 3 | ⟨100⟩ | 2.89 | 277 | residual tangled structure |
| **Fe–10Cr** | 50 | 1 | ½⟨111⟩ | 2.33 | 134 | absorbed, small cluster left |
| **Fe–10Cr** | 50 | 2 | ⟨100⟩ | 2.15 | 93 | fully absorbed |
| **Fe–10Cr** | 50 | 3 | ⟨100⟩ | 2.23 | 81 | residual ⟨100⟩ loop |
| **Fe–10Cr** | 70 | 1 | ½⟨111⟩ | 2.77 | 468 | absorbed, small cluster left |
| **Fe–10Cr** | 70 | 2 | ⟨100⟩ | 2.45 | 97 | fully absorbed |
| **Fe–10Cr** | 70 | 3 | mixed ½⟨111⟩/⟨100⟩ | 2.62 | 106 | absorbed, small cluster left |
| **Fe–10Cr** | 100 | 1 | ⟨100⟩ | 2.83 | 168 | residual ⟨100⟩ loop |
| **Fe–10Cr** | 100 | 2 | ⟨100⟩ | 2.93 | 334 | residual ⟨100⟩ loop |
| **Fe–10Cr** | 100 | 3 | ⟨100⟩ | 2.85 | 131 | residual ⟨100⟩ loop |

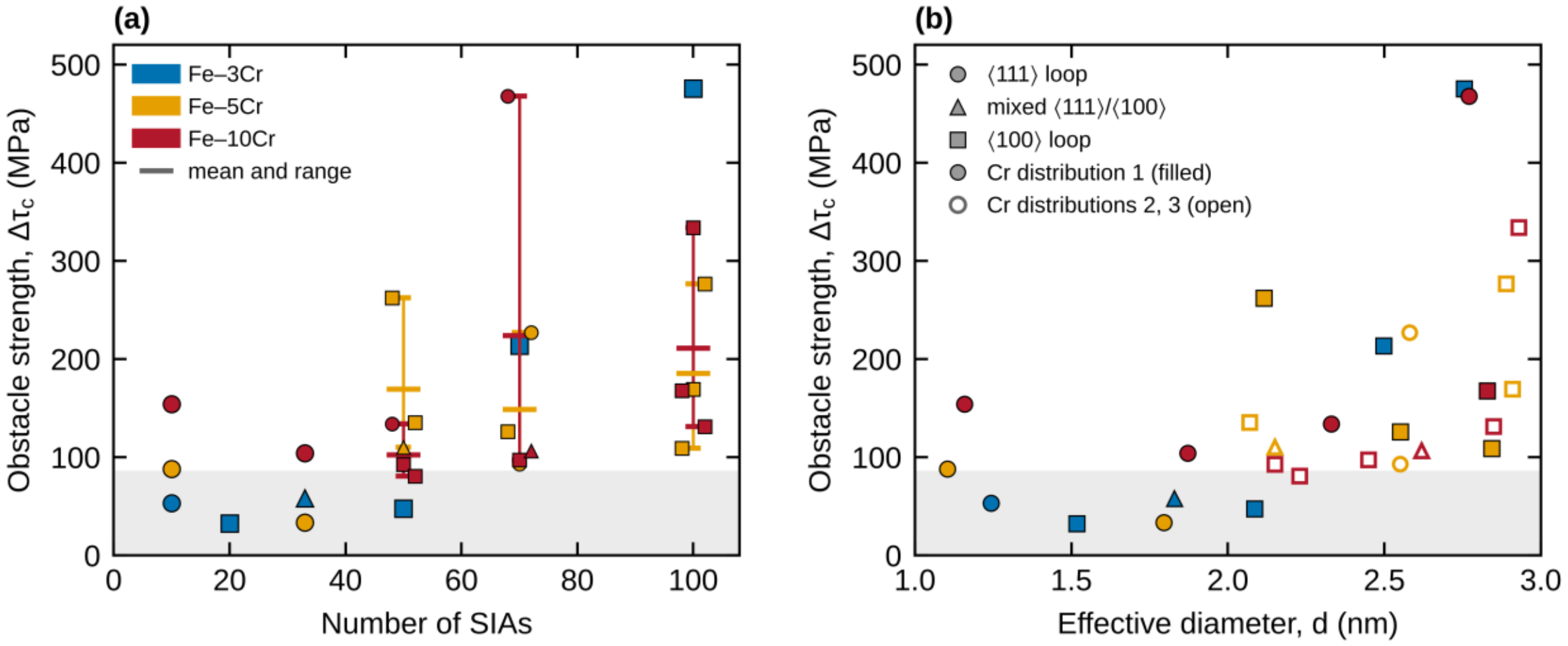

*Figure S1 – Obstacle strength of interstitial clusters for all Cr distributions. (a) Mean (bar) and range over three Cr distributions versus number of SIAs, with individual values (small symbols, offset for clarity). (b) Each simulation at its own effective diameter; filled symbols seed 1, open symbols seeds 2 and 3. Symbols indicate loop character (circle ½⟨111⟩, triangle mixed, square ⟨100⟩). The grey band marks the baseline fluctuation of the defect-free alloys.*

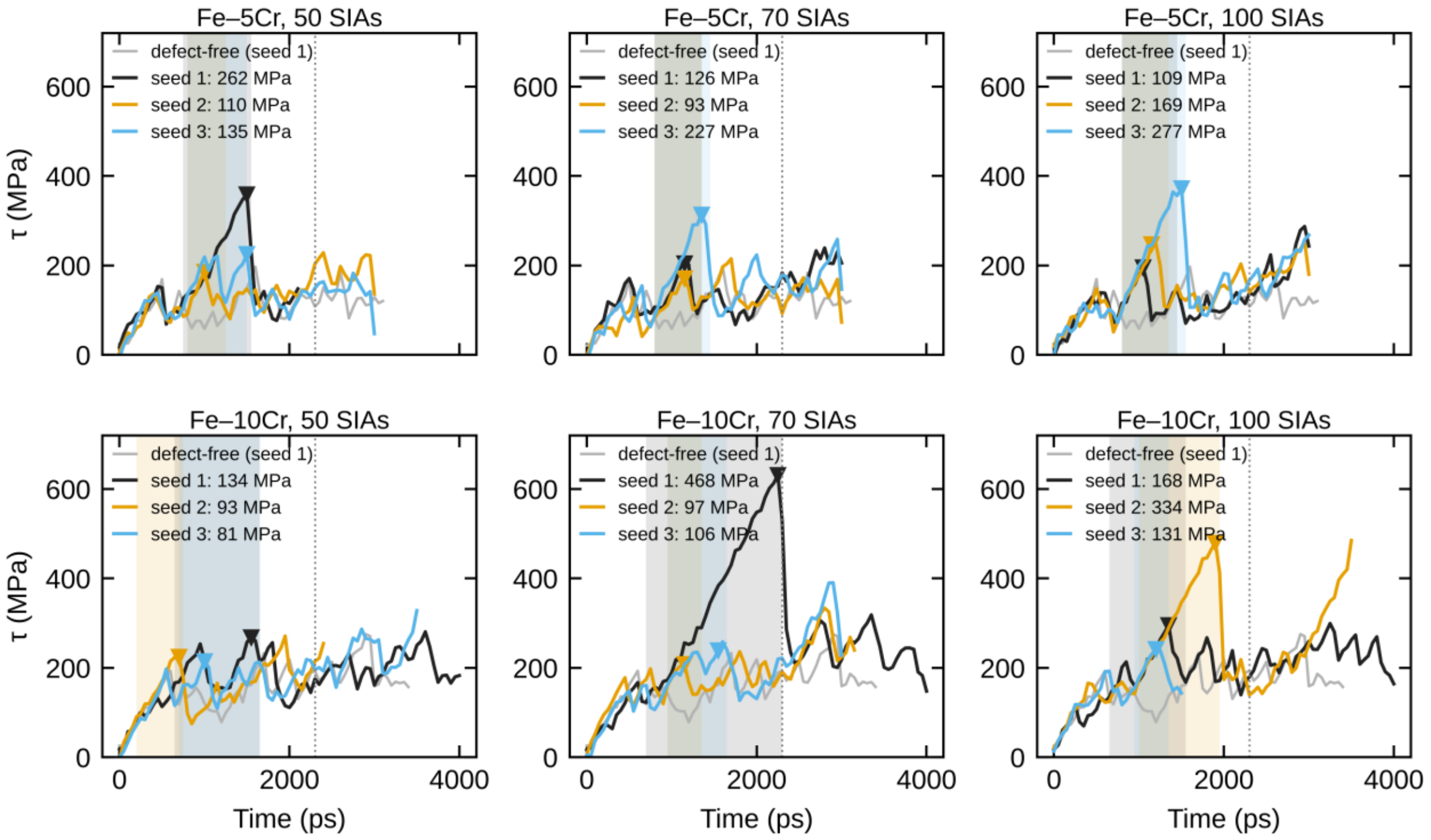

*Figure S2 – Resolved shear stress versus time for each Cr distribution for (a–c) Fe–5Cr with 50, 70 and 100 SIAs and (d–f) Fe–10Cr with 50, 70 and 100 SIAs, with the defect-free seed-1 curve (grey). Shaded regions mark the interaction window (up to 2300 ps) and triangles the peak for each seed; the dotted line marks the end of the first-encounter window (2300 ps), after which the dislocation re-encounters the defect through the periodic boundary.*

## S2. Relaxed structure of interstitial clusters for 15 Cr distributions

Interstitial clusters of 50 SIAs (Fe–3Cr and Fe–10Cr) and 25 SIAs (Fe–3Cr) were built with the procedure of Section 2.2 of the main text and relaxed for 15 independent Cr distributions in a fully periodic cell of about 205 × 203 × 202 Å$^3$ (~722,100 atoms) without a dislocation (energy minimisation, then 2 ns at 300 K, NVT). The loop character of the relaxed cluster was determined by DXA in OVITO. In Fe–10Cr, the mixed structures consist of segments of two ½⟨111⟩ variants joined by a ⟨100⟩ segment.

Table S2 – Loop character of relaxed 50-SIA clusters for 15 Cr distributions in each alloy.

| **Alloy** | **⟨100⟩ loop** | **½⟨111⟩ loop** | **Mixed** |
|---|---|---|---|
| **Fe–3Cr** | 11 | 0 | 4 |
| **Fe–10Cr** | 7 (seeds 2, 3, 5, 7, 8, 10, 11) | 4 (seeds 4, 6, 9, 12) | 4 (seeds 1, 13, 14, 15) |

In Fe–3Cr, the same test at 25 SIAs gave 9 ½⟨111⟩ loops, 3 ⟨100⟩ loops and 3 unfaulted clusters.